\documentclass[aps,prb,twocolumn,longbibliography]{revtex4-1}

\usepackage{graphicx}
\usepackage{dcolumn}
\usepackage{bm}
\usepackage{hyperref}
\usepackage{amsmath, amssymb}
\usepackage{mathtools}
\usepackage{color}
\usepackage[none]{hyphenat}
\usepackage{float}

\hypersetup{pdftex,
  breaklinks=true,
  colorlinks=true,
  urlcolor=blue,
  linkcolor=blue,
  citecolor=blue,
  }

\begin{document}

\title{Transition-Metal Substitutions in MOF-5: A Hybrid-DFT Study of Oxygen Electrocatalysis}

\author{Pedro H. Souza}

\author{Walter Orellana}
\email{worellana@unab.cl}

\affiliation{Departamento de F\'isica y Astronom\'ia, Facultad de Ciencias Exactas, Universidad Andres Bello, Sazi\'e 2212, Santiago 8370136, Chile}
        
\date{\today}
  
\begin{abstract}
Hybrid density functional theory calculations were used to investigate the stability, electronic properties, and catalytic activity of transition-metal 
impurities substituting Zn sites (M\(_{\rm Zn}\)) in the MOF-5 metal-organic framework [Zn\(_4\)O(BDC)\(_3\), BDC = 1,4-benzenedicarboxylate], 
with M = Sc, Ti, V, Cr, Mn, Fe, Co, Ni, and Cu. Most substitutions introduce \(3d\) states within the MOF-5 band gap and induce localized magnetic 
moments. Partially occupied \(3d\) states promote orbital hybridization and charge transfer with oxygenated intermediates, activating the dopant
sites toward oxygen electrocatalysis. The oxygen reduction and evolution reaction (ORR and OER) activities were evaluated using the computational 
hydrogen electrode model. Mn\(_{\rm Zn}\) and Fe\(_{\rm Zn}\) exhibit calculated ORR overpotentials of 0.57~V, while Mn\(_{\rm Zn}\) 
and Co\(_{\rm Zn}\) yield OER overpotentials of 0.42 and 0.51~V, respectively, approaching those of benchmark precious-metal electrocatalysts. 
These findings identify Mn substitution as a promising strategy for achieving bifunctional ORR/OER activity in MOF-5 and establish a link between 
dopant electronic structure and catalytic performance to guide the design of MOF-based electrocatalysts.
\end{abstract}
 
\maketitle

\section{Introduction}
Metal-organic frameworks (MOFs) are crystalline porous materials formed through the coordination of inorganic metal nodes with organic 
linkers. Their high surface areas, well-defined pore architectures, and chemical tunability have motivated applications in gas storage and 
separation, sensing, photocatalysis, and electrocatalysis. Among them, MOF-5 [Zn$_4$O(BDC)$_3$ (BDC = 1,4--benzenedicarboxylate)], 
an early landmark MOF,\cite{li1999} is regarded as an archetypal framework and has been extensively employed as a model system for 
investigating structure-property relationships in porous coordination networks.\cite{gangu2022}

MOF-5 can be prepared by conventional solvothermal methods\cite{li2009,colin2016,shirin2019,jamali2023} and alternative microwave-assisted
sonochemical,\cite{yoo2009,cory2020,yi2024} and mechanochemical routes.\cite{daofei2017,barbara2020,sylwia2021} Its porous structure 
and structural versatility have led to applications in gas storage and separation,\cite{palla2020,zhu2023,risnita2023,kong2026} chemical sensing,
\cite{lv2018,guselnikova2019,pathinti2026} and drug delivery.\cite{alneghery2025,chen2019} However, pristine MOF-5 has a large electronic band 
gap, poor intrinsic conductivity, and limited redox activity, which restrict its applicability in processes requiring efficient charge transport, strong 
adsorbate interactions, and catalytic activation. 

Several strategies have been proposed to tailor the electronic structure and chemical reactivity of MOF-5, including linker functionalization,
\cite{pim2022,burigana2025} defect engineering,\cite{chen2024,he2025} guest incorporation, and metal substitution. Transition-metal substitution 
is particularly attractive because it modifies the local electronic environment of the Zn$_4$O clusters while preserving the framework topology. 
Theoretical results have shown that partially filled $3d$ orbitals can introduce localized states near or within the band gap, alter the electronic 
structure around the Fermi level, generate magnetic moments, and create active adsorption sites.\cite{orellana2024}  

Experimental studies have demonstrated the controlled synthesis of Ni-doped MOF-5 through partial substitution of Zn by Ni, resulting in enhanced 
catalytic activity for the reduction of NO by CO.\cite{he2022}Ni-doped MOF-5 has also shown improved water-capture performance under low-humidity 
conditions.\cite{aziz2025} Furthermore, combined experimental and theoretical studies have indicated that metal-doped MOF-5 with Mn, Fe, Co, 
or Ni retains stability comparable to that of the pristine framework, although Ni-doped MOF-5 exhibits comparatively lower thermal stability. Thermal 
decomposition generally begins above 450~$^{\circ}$C, defining the approximate upper limit of stability.\cite{panda2022} Substitution with Mn, 
Fe, or Co narrows the band gap and promotes ligand-to-metal charge transfer, thereby broadening the potential applications of MOF-5 in 
optoelectronic devices, chemical sensing, and photocatalysis. Likewise, substitution with Cu, Ag, or Au strengthens metal-ligand orbital coupling, 
enhances charge mobility, and reduces the band gap.\cite{yan2023}

The electronic inertness of pristine MOF-5 contrasts with the diverse electronic configurations enabled by transition-metal incorporation. 
Localized spin-polarized $3d$ states can enhance molecular adsorption and facilitate charge transfer between adsorbates and the 
framework.\cite{orellana2024} Metal substitution may therefore transform MOF-5 from a porous adsorbent into a multifunctional material 
for sensing, photocatalysis, and electrocatalysis by introducing catalytically active sites. 

Herein, we employ density functional theory (DFT) calculations to systematically investigate metal-doped MOF-5 for oxygen electrocatalysis. 
We examine the substitution of Zn by \(3d\) transition metals and determine how these dopants affect the structural, electronic, and magnetic 
properties of the framework. We further evaluate the catalytic activity of the substituted metal centers (M\(_{\rm Zn}\)) toward the ORR and OER. 
Our results suggest that incorporating \(3d\) dopants is an effective strategy for transforming the intrinsically inactive Zn\(_4\)O centers of 
MOF-5 into potentially active electrocatalytic sites. In particular, Mn, Fe, and Co emerge as the most promising dopants, with their partially 
occupied \(3d\) states and sizable local magnetic moments associated with a favorable balance in the adsorption energetics of oxygenated 
intermediates.

\section{Computational details}
Spin-polarized DFT calculations were performed using the Vienna Ab initio Simulation Package (VASP 6.5).\cite{vasp1,vasp2} 
Geometry optimizations and electronic-structure calculations were carried out with the screened hybrid Heyd--Scuseria--Ernzerhof (HSE06) 
functional, which incorporates 25\% exact exchange.\cite{HSE03,HSE06} The plane-wave basis set was expanded using an energy 
cutoff of 400~eV. The Brillouin zone was sampled using a 2$\times$2$\times$2 k-point grid,\cite{kpoints} for electronic-property calculations, 
whereas adsorption energies were calculated using only the $\Gamma$ point owing to the high computational cost of denser k-point 
sampling. Table~S1 of the Supplementary Information compares the O$_2$ adsorption energies on the Mn dopant obtained 
with the two sampling schemes, showing difference of 0.04~eV. 
Long-range van der Waals interactions were accounted for using the vdW-DF2 functional.\cite{DF2} All atomic positions and lattice parameters 
were fully optimized until the residual forces on each atom were below 0.025~eV/\AA. Electronic self-consistency was achieved using an 
energy convergence threshold of 10$^{-5}$~eV. 

The MOF-5 structure was modeled using a reduced 53-atom unit cell, corresponding to one octant of the  primitive 424-atom unit cell. The 
electronic properties of pristine and metal-doped MOF-5 calculated using the reduced unit cell agree well with those obtained using the 
primitive unit cell, as shown in Figures~S1 and S2. Convergence with respect to the plane-wave energy cutoff and k-point 
density was verified to ensure the reliability of the chosen computational parameters, as shown in Figure~S3. Data processing 
and structural visualization were performed using the VASPKIT\cite{vaspkit} and VESTA\cite{vesta} software packages, respectively.

The formation energy of a neutral M dopant substituting a Zn atom in MOF-5, denoted by $E_f({\rm M_{Zn}})$, was calculated as
\begin{equation}
E_f({\rm M_{Zn}}) = E_t({\rm M_{Zn}}) - E_t + \mu_{\rm Zn} - \mu_{\rm M},
\end{equation}
where $E_t$ and $E_t({\rm M_{Zn}})$ are the total energies of pristine MOF-5 and MOF-5 containing the M dopant, respectively. The 
quantities $\mu_{\rm Zn}$ and $\mu_{\rm M}$ represent the chemical potentials of Zn and M atoms evaluated from their corresponding 
bulk phases. A more negative formation energy indicates more energetically favorable dopant incorporation relative to pristine MOF-5 
and the chosen elemental reference states.

The ORR and OER electrocatalytic performance of the dopant M$_{\rm Zn}$ sites in MOF-5 was evaluated using thermodynamic 
overpotentials calculated within the computational hydrogen electrode (CHE) model introduced by N{\o}rskov and co-workers~\cite{norskov2004}. 
Within this formalism, overpotentials are derived from the adsorption free energies of the key oxygenated intermediates (OOH$^*$, O$^*$, 
and OH$^*$), calculated as
\begin{equation}
\Delta G = \Delta E + \Delta E_{\rm ZPE} - T \Delta S + \Delta E_{\rm solv},
\end{equation}
where $\Delta E$, $\Delta E_{\rm ZPE}$, and $T \Delta S$ denote the adsorption energy, zero-point-energy correction, and vibrational entropy 
contribution, respectively, while $\Delta E_{\rm solv}$ is the solvation correction. Solvent effects on adsorbate binding energies were evaluated 
separately for each metal-doped MOF-5 structure using the implicit continuum dielectric model implemented in VASPsol.\cite{mathew2014} This 
approach accounts for the dielectric response of the surrounding solvent and its influence on OH$^*$ and OOH$^*$ adsorption energetics, 
without explicitly describing hydrogen bonding with individual H$_2$O molecules. The calculated solvation corrections for all metal dopants
considered are reported in Table~S2. The $\Delta E_{\rm ZPE}$ and $T \Delta S$ terms were obtained from vibrational frequency 
calculations for the adsorbed intermediates. Only adsorbate vibrational modes were included, with the framework atoms held fixed. Further 
details of the reaction free-energy calculations are provided in the Supplementary Information.

\section{Results and Discussion}
\subsection{Structural and Electronic Properties}

\begin{figure}
\centering 
\includegraphics[width=7cm]{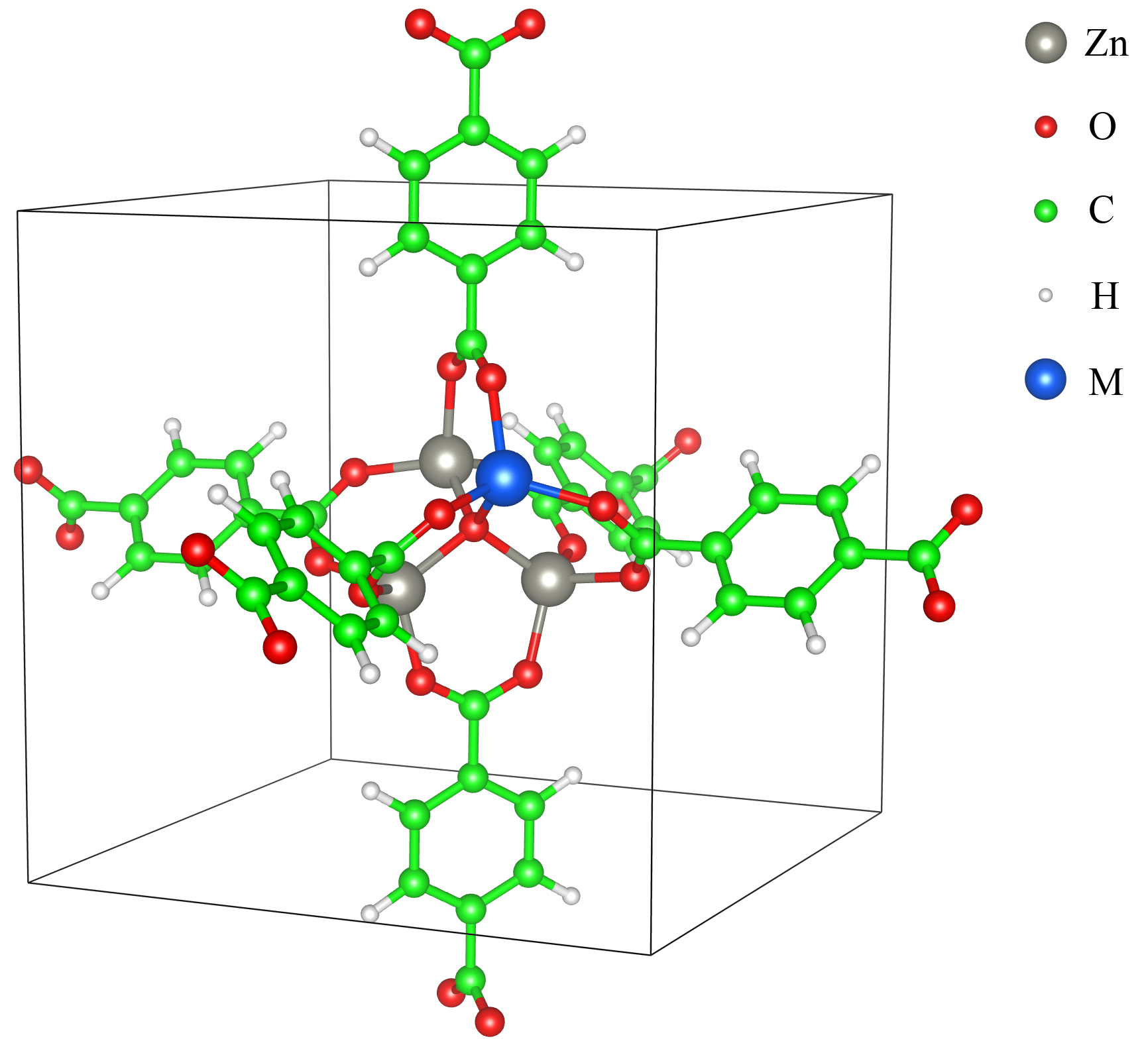}
\caption{Equilibrium geometry of metal-doped MOF-5 in the reduced unit-cell representation. M represents the metal substitution at a Zn site
(M$_{\rm Zn}$).}
\label{f1}
\end{figure}
Figure~\ref{f1} shows the atomic structure of metal-doped MOF-5, modeled using a reduced unit cell in which a metal dopant replaces one Zn 
atom in the tetrahedral Zn$_4$O cluster. The BDC ligands adopt a symmetric arrangement in the reduced cell, whereas their configuration is 
asymmetric in the primitive unit cell [see Figures~S1(a) and S1(b)]. Despite this structural difference, the two models yield qualitatively 
similar electronic properties for pristine and metal-doped MOF-5. The equilibrium lattice constant of the reduced cell is $a = 12.78$~\AA. Each Zn 
atom is coordinated to four O atoms, with an average Zn--O distance of 1.93~\AA, while the average nearest-neighbor Zn--Zn distance is 3.18~\AA.
\begin{table}[t]
\caption{Equilibrium geometries, magnetization, and energetics of the ${\rm M_{Zn}}$ impurity in MOF-5. $d_{\rm M-O}$ and $d_{\rm M-Zn}$ 
denote the average nearest-neighbor M--O and M--Zn distances, respectively (in \AA) and $m$ is the total magnetic moment (in $\mu_{\rm B}$). 
$E_{gap}$ and $E_f$ are the band gap and impurity formation energies (in eV), respectively. Results for pristine MOF-5 (M = Zn) are included for 
comparison.}
\begin{ruledtabular}
\begin{tabular}{lccccc}
M$_{\rm Zn}$ & $d_{\rm M-O}$ &  $d_{\rm M-Zn}$  & $m$ & $E_{gap}$ & $E_{f}$\\
\hline
Sc   & 2.03 & 3.24 & 1.0 & 0.91 & $-4.06$\\
Ti    & 2.06 & 3.26 & 2.0 & 1.73 & $-2.58$\\
V     & 2.02 & 3.27 & 3.0 & 2.25 & $-2.52$\\
Cr    &2.02 & 3.26 & 4.0 & 3.03  & $-2.31$\\
Mn   & 1.99 & 3.25& 5.0 & 4.38  & $-1.50$\\
Fe    & 1.99 &3.22 & 4.0  & 3.18 & $-0.84$\\
Co    & 1.93 &3.18 & 3.0 & 4.60 & $-0.35$\\
Ni     & 1.94 &3.24 & 2.0 & 4.51 & 0.16\\
Cu    & 1.95 & 3.19 & 1.0 & 2.95 & 1.70\\
Zn    & 1.94 & 3.18 & 0.0 & 5.02 & 0.0 \\
\end{tabular}
\end{ruledtabular}
\label{t1} 
\end{table}

Table~\ref{t1} summarizes the equilibrium geometries, magnetization, and energetics of substitutional M\(_{\rm Zn}\) dopants in MOF-5. 
Formation-energy calculations indicate that substitution is energetically favorable for dopants ranging from Sc to Co. The formation energy becomes progressively more negative from \(-0.35\)~eV for Co to \(-4.06\)~eV for Sc, suggesting that dopants with lower \(3d\)-shell occupations tend to 
exhibit more favorable incorporation energetics. Ni exhibits a slightly positive formation energy of 0.16~eV, indicating marginally unfavorable 
incorporation relative to the chosen reference states. Its comparatively less favorable incorporation is consistent with the experimental 
findings.\cite{panda2022} In contrast, Cu exhibits a substantially higher formation energy of 1.70~eV, suggesting energetically unfavorable 
incorporation. The total magnetic moment increases from $1~\mu_{\rm B}$ for Sc to a maximum of $5~\mu_{\rm B}$ for Mn, then decreases 
with increasing $3d$ occupation, reaching $1~\mu_{\rm B}$ for Cu. The spin-density distributions shown in Figure~S4 reveal that the 
magnetization is primarily localized on the metal dopant, except for Sc, for which spin polarization occurs mainly on the neighboring ligands 
[see Figure~S4(a)]. Metal substitution also narrows the band gap by introducing dopant-induced electronic states within the gap of pristine MOF-5.
\begin{figure*}
\centering 
\includegraphics[width=17cm]{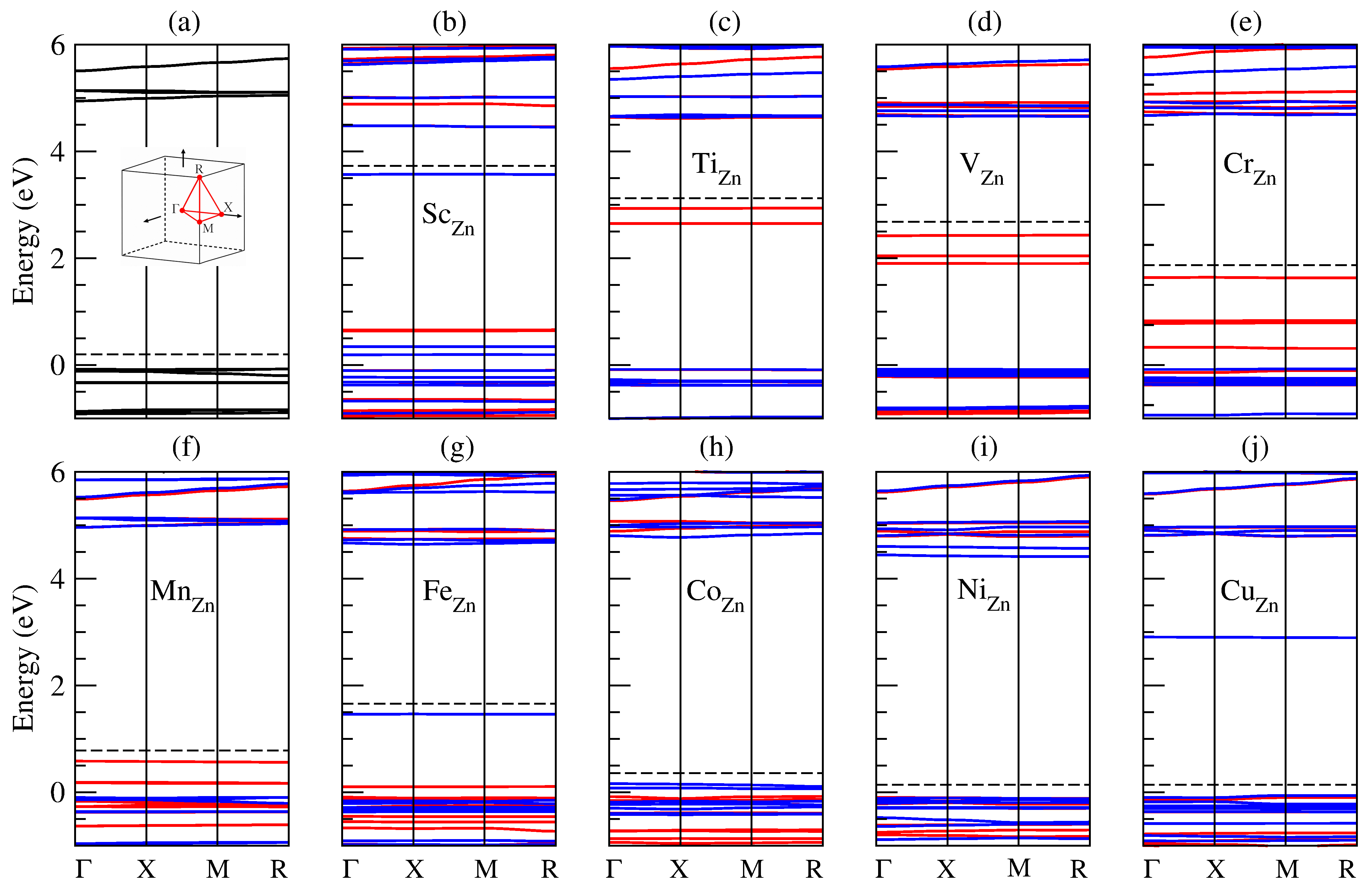}
\caption{Band structure calculations of metal-doped MOF-5: (a) pristine MOF-5, (b) MOF-5:Sc$_{\rm Zn}$, (c) MOF-5:Ti$_{\rm Zn}$, 
(d) MOF-5:V$_{\rm Zn}$, (e) MOF-5:Cr$_{\rm Zn}$, (f) MOF-5:Mn$_{\rm Zn}$,  (g) MOF-5:Fe$_{\rm Zn}$,  (h) MOF-5:Co$_{\rm Zn}$, 
(i) MOF-5:Ni$_{\rm Zn}$, and (j) MOF-5:Cu$_{\rm Zn}$. Red and blue lines represent spin-up and spin-down bands. The dashed line 
indicates the Fermi energy. The valence band maximum is set at zero energy.}
\label{f2}
\end{figure*}

Figure~\ref{f2} presents the calculated band structures of pristine and metal-doped MOF-5. Pristine MOF-5 exhibits a band gap of 5.0~eV, 
consistent with previous reports.\cite{orellana2024,souza2026,lee2025} The spin-polarized partial density of states (p-DOS), shown in 
Figure~S5, reveals that the valence band maximum (VBM) and conduction band minimum (CBM) of pristine MOF-5 are primarily 
derived from C and O orbitals, with a minor contribution from Zn. Metal substitution introduces states within the band gap, whose occupation 
and orbital character depend on the dopant. Sc introduces an occupied, predominantly ligand-derived state, characterized by a substantial C 
contribution and negligible Sc contribution. In contrast, the dopant-induced states in the remaining systems are dominated by metal orbitals, 
with smaller contributions from the surrounding framework. Ti, V, and Cr introduce multiple occupied states within the band gap, whereas the 
Mn- and Fe-induced states lie closer to the VBM. Co and Ni introduce unoccupied dopant states near the CBM, while the corresponding 
Cu-induced state lies near midgap.

The dopant-induced states may provide pathways for charge transfer and orbital hybridization with adsorbed molecules. Depending on their 
relative energy alignment and orbital symmetry, occupied metal-derived states may donate electron density to molecular acceptor orbitals, 
whereas unoccupied states may accept electron density from the adsorbate. In the case of O\(_2\), electron transfer into its antibonding \(\pi^*\) 
orbitals may weaken the O--O bond and promote molecular activation. The predominantly ligand-derived in-gap state in Sc-doped MOF-5 further 
suggests that the surrounding framework may also participate in adsorbate interactions. Collectively, these electronic features provide a qualitative 
basis for interpreting the adsorption behavior discussed below.

\subsection{ORR and OER Catalytic Activity}

\begin{table}[t]
\caption{Equilibrium geometry, magnetization, and energetics of the O$_2$ molecule adsorbed on the ${\rm M_{Zn}}$ impurity in MOF-5. 
$d_{\rm M-O}$ and $d_{\rm O-O}$ are atomic distances for the adsorbed O$_2$ and $m$ is the total magnetic moment (in $\mu_{\rm B}$). 
$E_b$ is the O$_2$ adsorption energy (in eV). Results for pristine MOF-5 (M = Zn) are included for comparison. The M--O$_2$ interactions 
are classified as chemisorption (chem) or physisorption (phys).}
\begin{ruledtabular}
\begin{tabular}{lccccc}
M$_{\rm Zn}$ & $d_{\rm M-O}$ & $d_{\rm O-O}$ & $m$ & $E_{b}$ & Interaction\\
\hline
Sc     & 2.11 & 1.33 & 1.0 & $-2.68$ &chem\\
Ti      & 1.86 & 1.46 & 0.0 & $-3.49$ &chem\\
V      & 2.00 & 1.33 & 1.0 & $-1.94$ &chem\\
Cr     & 2.00 & 1.32 & 2.0 & $-1.66$ &chem\\
Mn    & 2.04 & 1.29 & 3.0 & $-0.73$ &chem\\
Fe     & 2.02 & 1.26 & 6.0 & $-0.35$ &chem\\
Co     & 3.26 & 1.22 & 5.0 & $-0.13$ &phys\\
Ni      & 3.08 & 1.22 & 4.0 & $-0.12$ &phys\\
Cu    & 3.29 & 1.22 & 3.0 & $-0.12$ &phys\\
Zn     & 4.14 & 1.22 & 2.0 & $-0.15$ &phys\\
\end{tabular}
\end{ruledtabular}
\label{t2} 
\end{table}
The interaction of O$_2$ with the metal dopant site provides insight into molecular binding and activation in metal-doped MOFs, 
with implications for ORR initiation. Table~\ref{t2} summarizes the calculated adsorption energies, total magnetic moments, and 
equilibrium configurations of the O$_2$-adsorbed systems. For Co, Ni, and Cu impurities, O$_2$ binds weakly, with adsorption 
energies of approximately $-0.1$~eV and molecule--metal separations exceeding 3~\AA. These results are consistent with 
physisorption and suggest limited O$_2$ activation, although they do not exclude an ORR pathway involving OOH* formation 
without a stable O$_2$* precursor. For Fe and Mn, O$_2$ adsorption is moderately exothermic, with adsorption energies ranging 
from approximately $-0.4$ to $-0.7$~eV. In contrast, Cr, V, Ti, and Sc exhibit substantially stronger binding, with adsorption energies 
ranging from $-1.7$~eV for Cr to $-3.5$~eV for Ti. For these six impurities, the molecule--metal separations are approximately 
2~\AA, consistent with chemisorption, and O$_2$ adopts a side-on configuration with both oxygen atoms bonded to the metal center.

The O--O bond length increases from approximately 1.22~\AA\ for physisorbed O$_2$ to 1.26--1.46~\AA\ for chemisorbed O$_2$, 
providing structural evidence of bond weakening and molecular activation. However, O$_2$ activation alone does not ensure efficient 
catalysis. According to the Sabatier principle, an intermediate adsorption strength is required to balance molecular activation with the 
energetics of subsequent reaction steps. Moderately exothermic adsorption may therefore be more favorable than excessively strong 
binding, although this principle alone is insufficient to define an optimal adsorption-energy range.

\begin{figure}[t]
\centering 
\includegraphics[width=7cm]{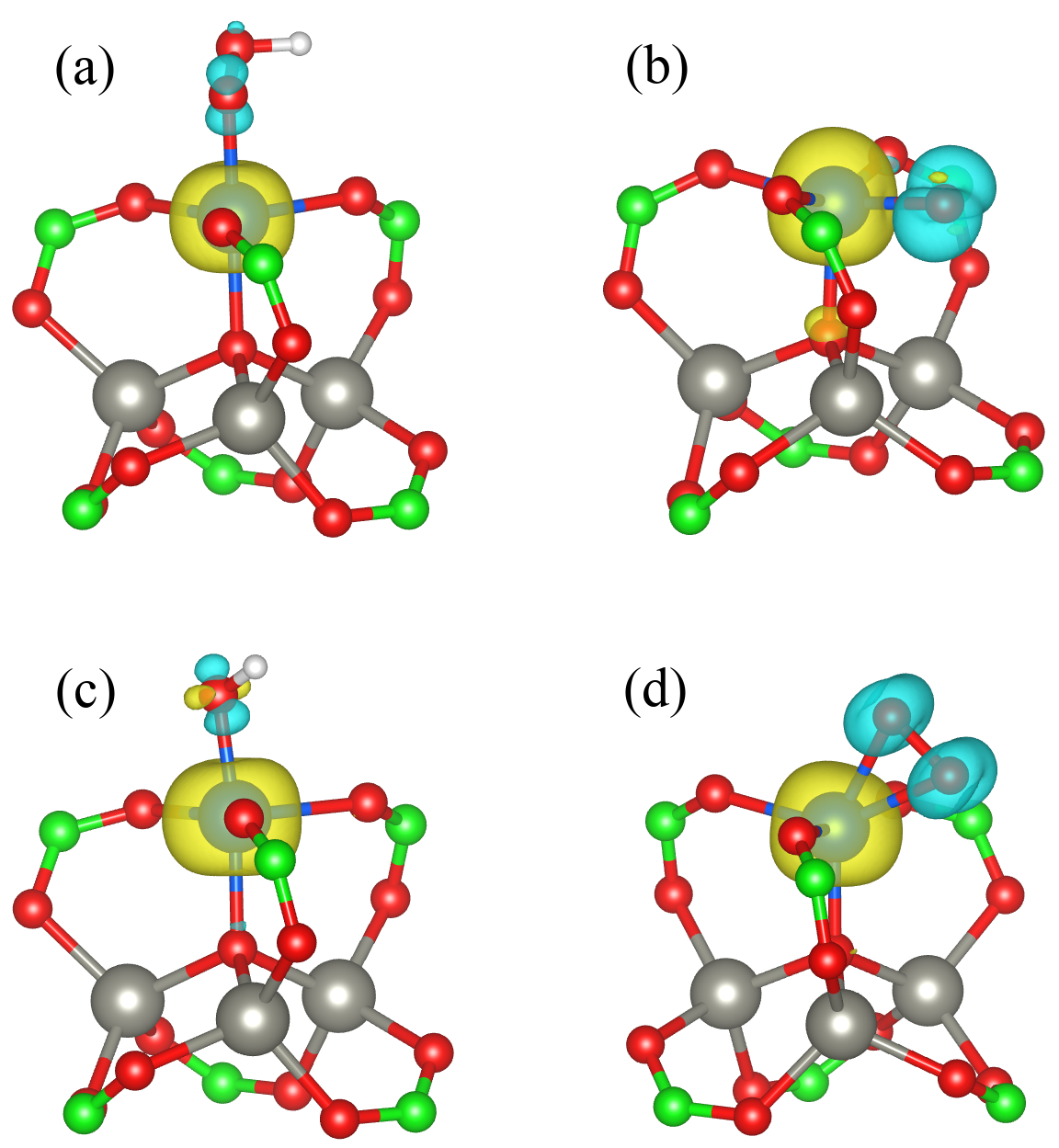}
\caption{Spin-density distribution and equilibrium geometry of ORR intermediates adsorbed on the Mn-doped MOF-5. (a) OOH$^*$, 
(b) O$^*$, and (c) OH$^*$. (d) The adsorbed O$_2$ in the side-on geometry is also shown for comparison. Yellow and cyan 
distributions represent spin-up and spin-down densities. The isosurface level corresponds to $\rho=0.015~a_{0}^{-3}$.}
\label{f3}
\end{figure}

The catalytic performance of the metal dopant site was evaluated within the framework of the CHE model considering the four-electron 
reaction pathway.\cite{norskov2004} Under alkaline conditions, hydroxide ions are the final reaction products, while water acts as the 
proton source. The overall oxygen reduction reaction can therefore be expressed as 
${\rm O}_{2} + 2\,{\rm H_{2}O} + 4 e^{-}  \longrightarrow 4\,{\rm OH}^{-}$.\cite{shilongli2024} Accordingly, the ORR is described by the 
following sequence of four elementary proton-electron transfer steps:

\begin{equation}
{\rm O}_{2} + {\rm H_{2}O} + e^{-} + * \leftrightharpoons {\rm OOH}^{*} + {\rm OH}^{-},
\end{equation}
\begin{equation}
{\rm OOH}^{*} + e^{-}  \leftrightharpoons {\rm O}^{*} + {\rm OH}^{-},
\end{equation}
\begin{equation}
{\rm O}^{*} + {\rm H_{2}O} + e^{-} \leftrightharpoons {\rm OH}^{*} +  {\rm OH}^{-},
\end{equation}
\begin{equation}
{\rm OH}^{*} + e^{-}  \leftrightharpoons * + {\rm OH}^{-},
\end{equation}

where \(*\) denotes the dopant adsorption site. The OER proceeds through the same elementary steps in 
the reverse direction. The reaction pathway involves three key oxygenated intermediates: OOH\(^{*}\), O\(^{*}\), and OH\(^{*}\). 
As a representative example, Figure~\ref{f3} shows their equilibrium geometries and spin-density distributions, together with 
those of adsorbed O\(_2^{*}\), on Mn-doped MOF-5.

To evaluate the catalytic behavior toward both the ORR and OER, adsorption free-energy diagrams were constructed for the 
key reaction intermediates. These profiles provide insight into the thermodynamic feasibility of the individual elementary steps 
and enable identification of the potential-limiting step. The reaction free energies were first evaluated at zero applied potential 
(\(U=0\)~V vs RHE). Because each elementary step involves the transfer of one electron, increasing the applied potential shifts 
its free energy by \(+eU\) for the ORR and by \(-eU\) for the reverse OER. At \(U=1.23\)~V, the overall four-electron ORR/OER 
is at thermodynamic equilibrium, although the individual elementary steps may remain either uphill or downhill depending on the 
adsorption free energies of the reaction intermediates.

The thermodynamic limiting potential or the onset potential (\(U_{\rm onset}\)) is defined as the highest applied potential at which 
all elementary ORR steps remain downhill in free energy or, for the OER, the lowest potential required to make all elementary steps 
downhill. This quantity provides the basis for evaluating catalytic performance through the theoretical overpotential (\(\eta\)), which 
measures the deviation of \(U_{\rm onset}\) from the thermodynamic equilibrium potential. Accordingly, the overpotential is calculated 
as \(\eta_{\rm ORR}=1.23~{\rm V}-U_{\rm onset}\) for the ORR and \(\eta_{\rm OER}=U_{\rm onset}-1.23~{\rm V}\) for the OER.
\cite{kulkarni2018} Further details of the methodology used to calculate the adsorption free energies, onset potentials, and 
overpotentials are provided in the Supplementary Information.

\begin{figure*}
\centering 
\includegraphics[width=14cm]{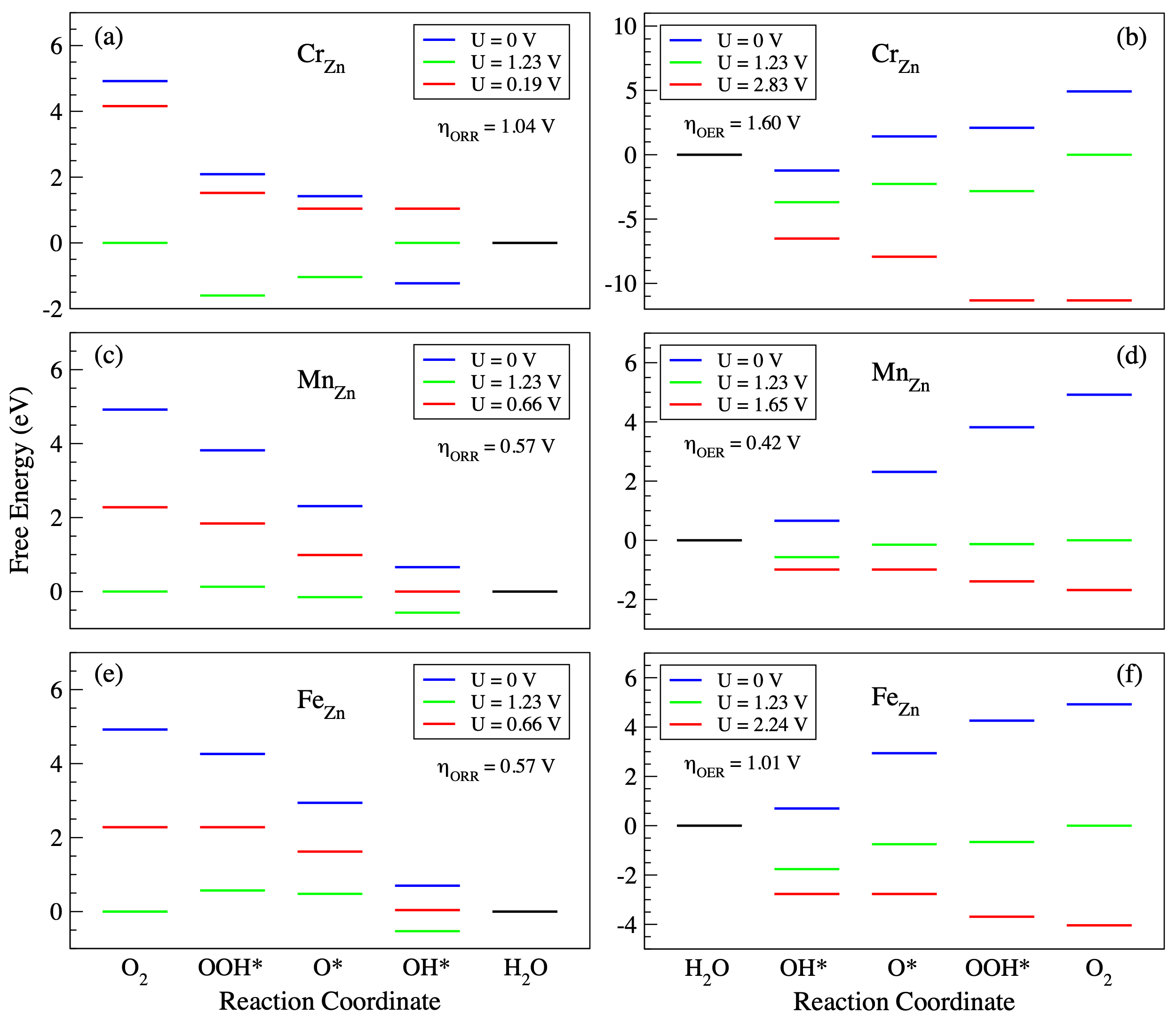}
\caption{Reaction free-energy diagrams for the four-electron ORR and the reverse OER pathway on (a,b) Cr${\rm Zn}$, (c,d) 
Mn${\rm Zn}$, and (e,f) Fe$_{\rm Zn}$ at the indicated potentials versus RHE. Blue, green, and red lines represent $U=0$~V, 
$U=1.23$~V, and the calculated limiting potential, respectively; $\eta$ denotes the corresponding overpotential.}
\label{f4}
\end{figure*} 

Figure~\ref{f4} presents the Gibbs free-energy profiles for the elementary ORR and OER steps on Cr-, Mn-, and Fe-doped 
MOF-5. The Cr dopant yields ORR and OER overpotentials of 1.04 and 1.60~V, respectively, which are substantially higher 
than the literature benchmark values of 0.43~V for the ORR on Pt(111)\cite{hansen2008} and 0.65~V for the OER on 
IrO\(_2\)(110),\cite{xu2024} obtained within comparable CHE frameworks. The poor performance of Cr-doped MOF-5 is 
consistent with the excessively strong binding of O\(_2\) and oxygenated intermediates at the Cr site. In contrast, Mn and Fe 
both exhibit lower ORR overpotentials of 0.57~V, approaching the Pt(111) benchmark. Mn and Co also show favorable OER 
performance, with overpotentials of 0.42 and 0.51~V, respectively, lower than the cited IrO\(_2\)(110) benchmark. These 
results identify Mn-doped MOF-5 as the most promising bifunctional ORR/OER catalyst among the systems investigated. 
Table~\ref{t3} summarizes the calculated onset potentials (\(U_{\rm onset}\)) and overpotentials (\(\eta\)) for both reactions 
across all metal substitutions. Sc, Ti, V, and Cr exhibit unfavorable reaction thermodynamics for both the ORR and OER, 
while Ni and Cu also display relatively high overpotentials. Fe and Co show reaction-selective behavior, favoring the ORR 
and OER, respectively. For comparison, pristine MOF-5 was evaluated by considering adsorption of the reaction intermediates 
at a Zn site. The resulting ORR and OER overpotentials of 2.26 and 1.27~V, respectively, indicate negligible intrinsic activity 
toward oxygen electrocatalysis.
\begin{table}[]
\caption{Theoretical results for ORR/OER onset potential ($U_{\rm onset}$ in V) and overpotential ($\eta$ in V) at the 
M$_{\rm Zn}$ active site in MOF-5. Available theoretical results for ORR on Pt(111)\cite{hansen2008} and OER on 
IrO$_2$(110)\cite{xu2024} are included as benchmark references. Results for pristine MOF-5 (M = Zn) are included for 
comparison.}
\label{t3}
\begin{ruledtabular}    
\begin{tabular}{lrccc}
 M$_{\rm Zn}$&
\multicolumn{2}{c}{ORR} &
\multicolumn{2}{c}{OER} \\
  \cline{2-3} \cline{4-5} 
 & $U_{\rm onset}$ & $\eta $ & $U_{\rm onset}$ & $\eta$  \\
 \hline
Sc         & $-4.66$   & 5.89 & 8.83 & 7.60\\
Ti           & $-4.14$   & 5.37 & 7.79 & 6.56\\
V           & $-0.18$   & 1.41 & 3.53 & 2.30\\
Cr          & 0.19        & 1.04 & 2.83 & 1.60 \\
Mn         & 0.66        & 0.57 & 1.65 & 0.42\\
Fe          & 0.66       & 0.57 & 2.24 & 1.01\\  
Co          & 0.23       & 1.00 & 1.74 & 0.51\\
Ni           & 0.11       &  1.12 & 2.31 & 1.08\\
Cu          & $0.02$   & 1.21 & 2.58 & 1.35\\
Zn          & $-1.03$  & 2.26  & 2.50 & 1.27\\ \\
Pt($111$)          & 0.80 & 0.43 & -- & --\\
IrO$_2$($110$) & -- & -- & 1.88 & 0.65 \\
\end{tabular}
\end{ruledtabular}
\end{table}

\begin{figure}[t!]
\centering 
\includegraphics[width=8cm]{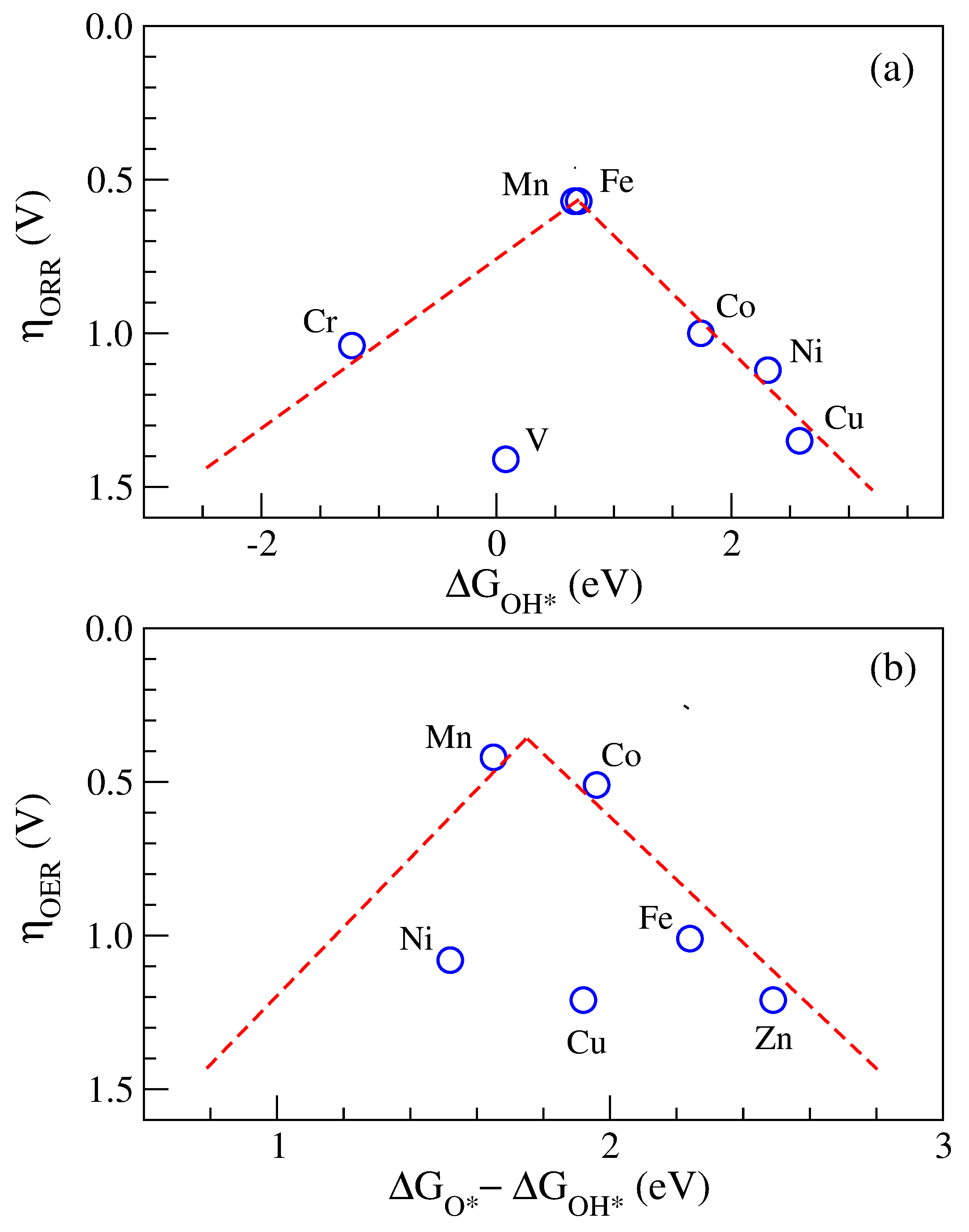}
\caption{Dependence of the calculated (a) ORR and (b) OER overpotentials on $\Delta G_{\rm OH^*}$ and 
$\Delta G_{\rm O^*}-\Delta G_{\rm OH^*}$ free-energy descriptors, respectively. The dashed lines are guides to the eye.}
\label{f5}
\end{figure}

To gain insight into the catalytic trends of metal-doped MOF-5, the calculated overpotentials were examined using established 
adsorption free-energy descriptors.\cite{man2011,ivanistsev2025} Figure~\ref{f5}(a) shows a volcano-like relationship between 
ORR performance and $\Delta G_{\rm OH^*}$, with Mn$_{\rm Zn}$ and Fe$_{\rm Zn}$ near the apex at approximately 0.7~eV. 
These sites exhibit the lowest ORR overpotentials among the systems shown, whereas deviations toward either stronger or weaker 
OH$^*$ binding generally correspond to poorer performance. V$_{\rm Zn}$ deviates substantially from this trend, suggesting that 
$\Delta G_{\rm OH^*}$ alone does not capture the energetic limitations associated with its strong oxygen affinity.
Figure~\ref{f5}(b) relates OER performance to $\Delta G_{\rm O^*}-\Delta G_{\rm OH^*}$. Mn$_{\rm Zn}$ and Co$_{\rm Zn}$ lie 
closest to the apex of the corresponding volcano, located at approximately 1.8~eV, with Mn$_{\rm Zn}$ exhibiting the lowest 
OER overpotential. However, the deviations of Ni$_{\rm Zn}$ and Cu$_{\rm Zn}$ indicate that a favorable value of this descriptor 
does not necessarily ensure low overpotentials, as other elementary steps may remain energetically demanding. Together, these 
results identify Mn substitution as the most promising option for bifunctional oxygen electrocatalysis among the systems compared.

Experimentally verified thermal stability\cite{panda2022} and favorable metal-substitution formation energies alone do not establish 
durability in alkaline electrolytes under ORR/OER operating potentials. Previous studies have reported water-induced structural changes 
and hydrolysis in MOF-5.\cite{rodriguez2015,ming2017} Our calculated formation energies indicate favorable Mn and Fe incorporation 
relative to the chosen reference states, while the overpotentials suggest promising bifunctional activity for Mn-doped MOF-5 and favorable 
ORR activity for Fe-doped MOF-5. However, these activity predictions describe the thermodynamic behavior of the modeled
frameworks within the CHE approach. Their practical relevance therefore remains contingent on framework stability under operating 
conditions, efficient charge transport, and favorable reaction kinetics.

Finally, we examine how the number of substitutional impurities within a single Zn\(_4\)O cluster influences the electrocatalytic activity 
of doped MOF-5. To this end, we compare the ORR and OER overpotentials of Zn\(_4\)O clusters containing either one or two metal 
impurities (M\(_1\)Zn\(_3\) and M\(_2\)Zn\(_2\), respectively), with M = Mn, Fe. Increasing the number of impurities from one to two 
changes the calculated overpotentials by less than 0.05~V for both reactions, as shown in Tables~S3 and S4. Within 
the configurations investigated, these small variations indicate that dopant occupancy within an individual Zn\(_4\)O cluster has little 
influence on the ORR and OER overpotentials of Mn- and Fe-doped MOF-5.\\

\section{Summary and Conclusions}
In summary, we have investigated the effects of transition-metal substitution on the electronic, magnetic, and oxygen electrocatalytic 
properties of MOF-5 using DFT calculations. Substitution of Sc, Ti, V, Cr, Mn, Fe, Co, Ni, and Cu at Zn sites 
preserves the main structural features of the framework while substantially modifying its electronic structure. Dopant-induced states 
narrow the band gap of pristine MOF-5 and give rise to spin-polarized electronic states and local magnetic moments, demonstrating 
that substitutional metal doping can effectively tailor the electronic and magnetic properties of the framework.

The catalytic analysis reveals that these electronic modifications strongly influence the adsorption of oxygenated intermediates and, 
consequently, the ORR and OER energetics. Compared with pristine MOF-5, dopants featuring partially occupied \(3d\) states and 
relatively large local magnetic moments reduce the theoretical overpotentials. Mn and Fe dopants exhibit the lowest calculated ORR 
overpotentials, both at 0.57~V, whereas Mn and Co show promising OER performance, with overpotentials of 0.42 and 0.51~V, 
respectively. In contrast, Sc, Ti, V, and Cr exhibit substantially higher overpotentials because of their strong affinity for oxygenated 
intermediates. Co, Ni, and Cu generally bind these species more weakly; however, Co achieves a favorable balance between the 
adsorption free energies of O\(^{*}\) and OH\(^{*}\), resulting in promising OER activity. These trends are consistent with the ORR 
and OER volcano plots relating overpotentical to the free-energy descriptors \(\Delta G_{\rm OH^*}\) and 
\(\Delta G_{\rm O^*}-\Delta G_{\rm OH^*}\), respectively.

Overall, our findings identify transition-metal substitution as a promising strategy for converting the Zn$_4$O centers of MOF-5 into
potentially active sites for oxygen electrocatalysis. Mn$_{\rm Zn}$ emerges as the most promising bifunctional site among the systems
investigated, combining favorable incorporation energetics with low calculated ORR and OER overpotentials. These results provide a
theoretical basis for tailoring the electronic and electrocatalytic properties of MOF-based materials and may guide experimental efforts
toward the development of low-cost single-atom electrocatalysts for energy conversion.

\begin{acknowledgments}
We acknowledge financial support from the National Agency for Research and Development of Chile (ANID), 
through projects FONDECYT 1230138 (W.O.) and FONDECYT POSTDOC 3240440 (P.H.S.). The computational 
resources were provided by Fenix HPC (UNAB) and the supercomputing infrastructure of the NLHPC (CCSS210001).
\end{acknowledgments}

\bibliography{mybib}
\end{document}